\documentclass{PoS}
\usepackage{tabularx}
\usepackage{graphicx}
\usepackage{amsmath} 
\usepackage{braket}
\usepackage{cleveref}
\usepackage{slashed}
\usepackage[T1]{fontenc}
\usepackage[utf8x]{inputenc}
\usepackage{booktabs}
\usepackage[rgb]{xcolor}
\usepackage{mathtools}

\definecolor{mypink}{rgb}{1.0 0.752941176471 0.796078431373}
\definecolor{mygray}{rgb}{0.501960784314 0.501960784314 0.501960784314}
\definecolor{myfirebrick}{rgb}{0.698039215686 0.133333333333 0.133333333333}
\definecolor{mygreen}{rgb}{0.0 0.501960784314 0.0}
\definecolor{myred}{rgb}{1.0 0.0 0.0}
\definecolor{myorange}{rgb}{1.0 0.647058823529 0.0}
\definecolor{mycyan}{rgb}{0.0 1.0 1.0}
\definecolor{mypurple}{rgb}{0.501960784314 0.0 0.501960784314}
\definecolor{mysalmon}{rgb}{0.980392156863 0.501960784314 0.447058823529}
\definecolor{myblue}{rgb}{0.0 0.0 1.0}
\definecolor{mybrown}{rgb}{0.647058823529 0.164705882353 0.164705882353}
\newcommand\crule[3][black]{\textcolor{#1}{\rule{#2}{#3}}}
\title{Nucleon generalized form factors from lattice QCD with nearly physical quark masses}
\ShortTitle{Nucleon generalized form factors} 

\author{Gunnar Bali,$^a$ Sara Collins,$^a$ Meinulf Göckeler,$^a$ \speaker{Rudolf Rödl},$^a$ 
  Andreas Schäfer,$^a$ Andre Sternbeck$^{b}$ \\
  \llap{$^a$} Institut für Theoretische Physik, Universität Regensburg, 93040 Regensburg, Germany\\
  \llap{$^b$} Theoretisch-Physikalisches Institut, Friedrich-Schiller-Universität Jena, 07743 Jena, Germany\\
  E-mail: \email{Rudolf.Roedl@physik.uni-regensburg.de}
}  

\abstract{
  We determine generalized form factors of the nucleon from lattice 
  simulations with $N_f = 2$ mass-degenerate non-perturbatively improved Wilson-Sheikholeslami-Wohlert 
  fermions down to a pion mass of 150 MeV. We also present the resulting isovector quark angular momentum. 
  Possible excited-state contaminations are investigated with correlated simultaneous fits.
}

\FullConference{The 33rd International Symposium on Lattice Field Theory\\
  14 -18 July 2015\\
  Kobe International Conference Center, Kobe, Japan*}

\begin{document}
\section{Introduction}
\label{sec-introduction}
Generalized Parton Distributions (GPDs) were introduced in the '90s
\cite{Mueller:1998fv, Ji:1996ek}.  They describe the inner structure
of the nucleon.  The starting point is the parametrization of the
off-diagonal matrix elements of the light cone operator
\cite{Ji:1998pc, Ji:1996nm, LHPC:2003aa, Diehl:2003ny}
\begin{align}
  \label{eq-decomp}
  \bra{N(P^\prime)}\mathcal{O}_q(x)\ket{N(P)} \overset{\mathrm{twist \ 2}  }{=}\frac{1}{2} 
  \overline{u}(P') \left[  
  H_q(x, \xi, t ) \ \slashed{n} +  
  E_q(x, \xi, t ) \, \frac{\mathrm{i} \sigma^{\mu \nu} n_\mu \Delta_\nu}{2m_N} 
  \right] u(P), \\  
\label{eq-lc}
  \mathcal{O}_q(x) =  \frac{1}{2} \int \limits_{-\infty}^{+\infty} \frac{\mathrm{d} \, 
  \lambda}{2\pi} \, \mathrm{e}^{\mathrm{i} \lambda x} \,
  \overline{\psi_q}\left(- \frac{\lambda}{2} n \right)
  \slashed{n} \, \mathcal{P} \exp 
  \left [  
  -\mathrm{i} \, g \int \limits_{-\frac{\lambda}{2}}^{+\frac{\lambda}{2}} \mathrm{d} \,
  \alpha \  A^\beta(\alpha \, n) \, n_\beta
  \right] 
  \psi_q\left(+ \frac{\lambda}{2} n \right).
\end{align}
The path ordering operator $\mathcal{P}$ in \cref{eq-lc} ensures gauge
invariance.  Further, it depends on the momentum fraction $x$ and the
quark flavor $q$.  The spinors $\overline{\psi}(\cdot)$ and
$\psi(\cdot)$ are parametrized along the light cone ($n^2 = 0$) by the parameter
$\lambda$.

Key ingredients of the parameterization of the matrix element in
\cref{eq-decomp} are the constraint given by the Lorentz structure and
the definitions of $H_q( \cdot) $ and $E_q(\cdot)$, which depend on
the momentum fraction $x$ and the kinematic variables
$\xi = -n^\mu \Delta_\mu/2$ and $t=\Delta^2$ with
$\Delta^\mu = P'^\mu - P^\mu$.

The direct calculation of GPDs by virtue of lattice QCD is not
possible since $\mathcal{O}_q(x)$ is a light cone operator.  However,
one can expand \cref{eq-lc} in towers of local twist two operators
\begin{align}
  \MoveEqLeft[3] \bra{N(P^\prime)} 
  \overline{\psi}_q \gamma^{\{\mu} \ 
  \mathrm{i} \overset{\leftrightarrow}{D}^{\mu_1}  \,  \cdots \,
  \mathrm{i} \overset{\leftrightarrow}{D}^{\mu_n \}} \, \psi_q \,
  \ket{N(P)} = \nonumber \\
&= \overline{u}(P') \left\{ \frac{\null}{\null} \right .
  \sum \limits_{i=0}^{n}  \delta_{i, \mathrm{even}} \,
  \gamma^{\{\mu}\, \Delta^{\mu_1}\cdots\Delta^{\mu_i} \,
  \overline{P}^{\mu_{i+1}}\cdots \overline{P}^{\mu_{n} \}} \
  \, A^q_{n+1, i}(t)  \nonumber  \\ 
& \quad+\frac{-\mathrm{i}}{2m}\,\sum\limits_{i=0}^{n}\delta_{i, \mathrm{even}} \,
  \Delta_{\alpha} \sigma^{\alpha \{\mu}  \,  
  \Delta^{\mu_1}\cdots\Delta^{\mu_i} \, \overline{P}^{\mu_{i+1}}\cdots \overline{P}^{\mu_{n} \} } \
  \, B^q_{n+1, i}(t)  \nonumber \\
&\quad + \frac{1 }{m}\,  \delta_{n, \mathrm{odd}}  \, \Delta^{\mu} \cdots   \Delta^{\mu_n} \,
  \, C^q_{n+1}(t) \left . \frac{\null}{\null}  \right\} u(P).
  \label{eg-gpd-gff}
\end{align}
The coefficient functions $A^q_{n+1, i}(t)$, $B^q_{n+1, i}(t)$ and
$C^q_{n+1}(t)$ are so called (vector) Generalized Form Factors (GFFs)
which are related to (vector) GPDs by
\begin{align}
\int_{-1}^{+1} \mathrm{d}\, x \ x^{n} \, H^q(x, \xi, t) &= 
\sum \limits_{i=0}^{n}  \delta_{i, \mathrm{even}} 
(-2\xi)^i A^q_{n+1, i}(t) +  \delta_{n, \mathrm{odd}} (-2\xi)^{n+1} \,  C^q_{n+1}(t), \\ 
\int_{-1}^{+1} \mathrm{d}\, x \ x^{n} \, E^q(x, \xi, t) &= 
\sum \limits_{i=0}^{n}  \delta_{i, \mathrm{even}} 
(-2\xi)^i B^q_{n+1, i}(t) -  \delta_{n, \mathrm{odd}} (-2\xi)^{n+1} \,  C^q_{n+1}(t).
\end{align}
The knowledge of $A^{u-d}_{2,0}(t)$, $B^{u-d}_{2,0}(t)$ is of
particular interest since they encode the total angular momentum
\begin{align}
  \label{eq-ji}
  J^{u-d} = \frac{1}{2} \left( A_{2,0}^{u-d}(t=0) + B_{2,0}^{u-d} (t=0) \right).
\end{align}
Moreover, we determine $\tilde{A}^{u-d}_{2,0}(t)$ and
$\tilde{B}^{u-d}_{2,0}(t)$ which are the GFFs corresponding to the
axial GPDs.

\section{Lattice set-up}
\label{sec-latt-set-up}
We use $N_f = 2$ mass-degenerate non-perturbatively improved
Wilson-Sheikholeslami-Wohlert fermions with Wilson glue.  The
necessary gauge configurations were produced by the QCDSF
collaboration and RQCD (Regensburg QCD).  The lattice spacing was set
as described in \cite{Bali:2012qs}.  To ensure ground-state dominance
we use optimized Wuppertal smearing as described in
\cite{Bali:2014nma}.  All results are computed at a renormalization
scale $\mu = 2 \, \mathrm{GeV}$ using non-perturbatively improved
conversion factors.  In \cref{latsetup} we give an ensemble overview.
\begin{table}[htp]
  \begin{center}
    \begin{tabular}{lcccccccc}
      Ensemble& $\beta$&$a$[fm]&$\kappa$ &$V$&$m_{\pi}$[GeV]&$Lm_{\pi}$&$N_{\mathrm{conf}}$&$t_{\mathrm{f}}/a$\\
      \hline
      \crule[mypink]{0.3cm}{0.3cm} I&5.20&0.081&0.13596&$32^3\times 64$ & 0.2795(18)&3.69&$1986(4)$&13\\
      \hline
      \crule[mygray]{0.3cm}{0.3cm} II&5.29&0.071&0.13620&$24^3\times 48$&0.4264(20)&3.71&$1999(2)$&15\\
      \crule[myfirebrick]{0.3cm}{0.3cm} III\hfill&&&0.13620&$32^3\times 64$&0.4222(13)&4.90&$1998(2)$&15,17 \\
 \crule[mygreen]{0.3cm}{0.3cm} IV&&&0.13632 &$32^3\times 64$&0.2946(14)&3.42&$2023(2)$&7(1),9(1),11(1) \\
      &&&&&&&&13,15,17\\
      \crule[myred]{0.3cm}{0.3cm} V&&&&$40^3\times 64$&0.2888(11)&4.19&$2025(2)$&15\\
      \crule[myorange]{0.3cm}{0.3cm}VI&&&&$64^3\times 64$&0.2895(07)&6.71&$1232(2)$&15 \\
      \crule[mycyan]{0.3cm}{0.3cm} VII&&&0.13640&$48^3\times 64$&0.1597(15)&2.78&$3442(2)$&15 \\
      \crule[mypurple]{0.3cm}{0.3cm} VIII&&&&$64^3\times 64$&0.1497(13)&3.47&$1593(3)$&9(1),12(2),15\\
      \hline
      \crule[mysalmon]{0.3cm}{0.3cm} IX&5.40 & 0.060 &0.13640&$32^3\times 64$&0.4897(17)&4.81 & $1123(2)$ &   17 \\
      \crule[myblue]{0.3cm}{0.3cm} X&&&0.13647&$32^3\times 64$&0.4262(20)&4.18&$1999(2)$&17\\
      \crule[mybrown]{0.3cm}{0.3cm} XI&&&0.13660&$48^3\times 64$   & 0.2595(09) &  3.82   &   $2177(2)$   &    17
    \end{tabular}
    \caption{$N_f = 2$ lattice set-up. Number of sources per
      configuration in brackets.}
    \label{latsetup}
  \end{center}
\end{table}
\section{Extracting Generalized Form Factors}
GFFs are obtained by solving an (in general) overdetermined system of
equations.  In the case of the vector GFFs the equation system reads
\begin{align}
  \epsilon(t, \tau_{\mathrm{sink}} ) = 
  \left[\mathrm{M} 
  \begin{bmatrix}
    A_{2,0}(t)\\
    B_{2,0}(t) \\
    C_{2}(t)\\
  \end{bmatrix} - \vec{c}(t, \tau_{\mathrm{sink}} ) \right]^T
\mathrm{cov}^{-1}(\vec{c}(t, \tau_{\mathrm{sink}} )) \left[\mathrm{M}
  \begin{bmatrix}
    A_{2,0}(t)\\
    B_{2,0}(t)\\
    C_{2}(t)\\
  \end{bmatrix} - \vec{c}(t, \tau_{\mathrm{sink}} ) \right].
  \label{eg-eqs-vec}
\end{align}
We extract the GFFs of interest by minimizing \cref{eg-eqs-vec}.  The
coefficient matrix $M$ in \cref{eg-eqs-vec} is fully determined by
\cref{eg-gpd-gff}.  The matrix elements
$\vec{c}(t, \tau_{\mathrm{sink}} )$ are extracted from lattice
three-point functions.  The dependency on the source-sink separation
$ \tau_{\mathrm{sink}}$\footnote{We set $\tau_{\mathrm{source}}=0.$}
is examined in the next section, where in the limit
$ \tau_{\mathrm{sink}} \rightarrow \infty$ the ground state GFFs are
obtained.

A good signal is a prerequisite for excited-state fits since we have
to fix many fit parameters.  It turns out that excited-state fits are
not possible for $t < 0$ if we naively implement \cref{eg-eqs-vec}.
However, we can dramatically improve the signal if we average
three-point functions which lead to the same rows in $M$.
\newpage

\section{Extraction of matrix elements from lattice QCD and excited
  states.}
As mentioned in \cref{sec-latt-set-up} we use optimized Wuppertal
smearing on our quark field interpolators.  Possible remaining excited
states contributions are treated with simultaneous combined fits to
two- and three-point functions. We therefore parametrize the two and
three-point functions
\begin{align}
  \label{fa}
  C_3(\vec{p}_i, \tau, \tau_{\mathrm{sink}}) &= c( \tau_{\mathrm{sink}})  \sqrt{z_N^{\vec{0}} \ z_{N}^{\vec{p}_i}} \ \mathrm{e}^{-m_N( \tau_{\mathrm{sink}} - \tau) } \mathrm{e}^{-E(\vec{p}_i, \ m_N) \tau} \nonumber \\
                                             &+ \ x_1 \  \ \mathrm{e}^{-m_N( \tau_{\mathrm{sink}} - \tau) } \mathrm{e}^{-E^\prime(\vec{p}_i) \tau}
                                               + \ x_2 \   \mathrm{e}^{-m_N^\prime( \tau_{\mathrm{sink}} - \tau) } \mathrm{e}^{-E(\vec{p}_i, \ m_N) \tau}
                                               + \ x_3  \  \mathrm{e}^{-m_N^\prime( \tau_{\mathrm{sink}} - \tau) } \mathrm{e}^{-E^\prime(\vec{p}_i) \tau}, \nonumber \\ \nonumber
  C_2^{1e}(p,\tau)&=z_N^{\vec{p}} \ \frac{E(\vec{p},\ m_N)+m_N}{E(\vec{p},\ m_N)} \ \mathrm{e}^{-E(\vec{p},\ m_N) \tau}, \nonumber \\ 
  C_2^{2e}(p,\tau)&=z_N^{\vec{p}} \ \frac{E(\vec{p},\ m_N) + m_N}{E(\vec{p},\ m_N)} \ \mathrm{e}^{-E(\vec{p}, \ m_N) \tau}  + z_{N^\prime}^{\vec{p}} \  \frac{E^\prime(\vec{p}) + m_{N^\prime} }{E^\prime(\vec{p})} \ \mathrm{e}^{-E^\prime(\vec{p}) \tau} .
\end{align}
Above we denote the nucleon mass as $m_N$ and its energy as
$E(\vec{p},m_N)$.  Further, we assume the continuum dispersion
relation for the ground state $E(\vec{p},m_N) = m_N^2 +\vec{p}^2$ .
All expressions corresponding to the first excited-state are indicated
by a prime.  The energy of the first excited-state $E^\prime(\vec{p})$
is left as a fit parameter since this could be a mulit-hadronic state.
Our kinematic set-up is chosen such that the final momentum is always
zero $ \vec{p}_f = \vec{0}$.  To accomplish the ground state
extraction (parameter $c( \tau_{\mathrm{sink}})$ ) we fix
$z_N^{\vec{0}}$ and $z_{N}^{\vec{p}_i}$ by virtue of the two-point
functions.

\section{Assessment of the excited-state fit}
\begin{figure}[!htpb]
  \centering
  \includegraphics[width=1.0\textwidth]{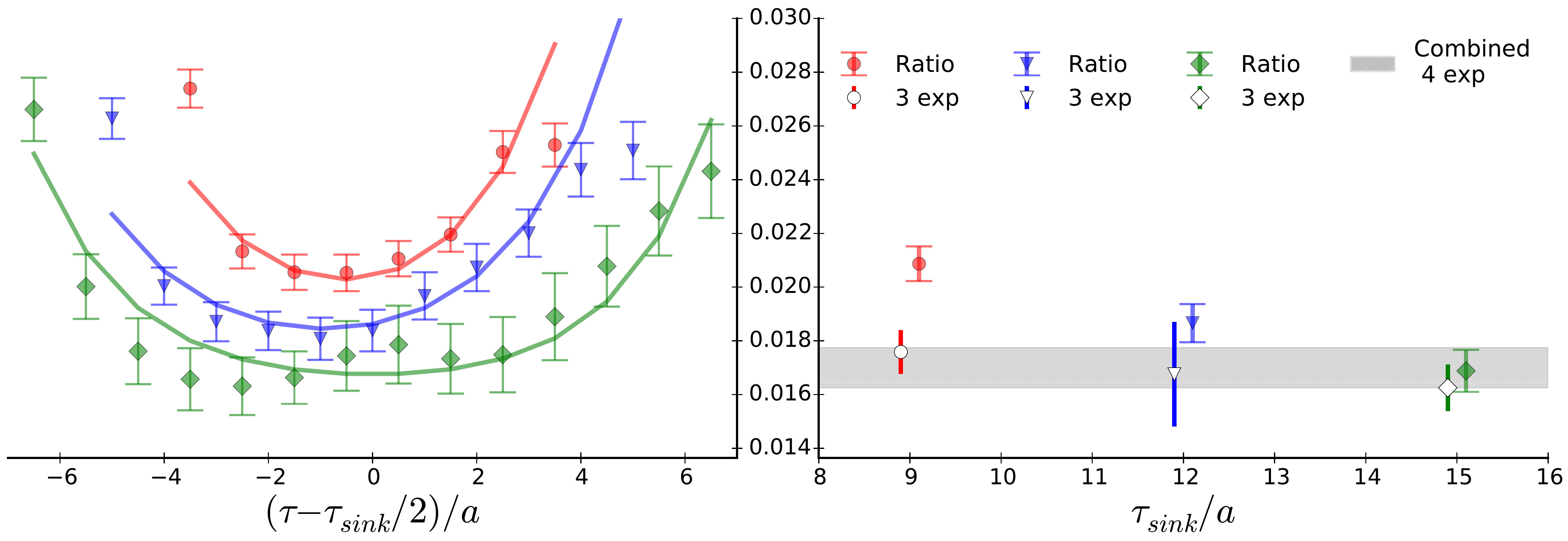}
  \caption{Extraction method analysis: Data points in the left panel
    are ratios of three-point and two-point functions for different
    source sink separations of ensemble VIII for a certain row $r_1$
    in the coefficient matrix $M$ at $t = -0.211 \, \mathrm{GeV}^2$.  The
    corresponding solid lines are ratios created with the fit
    functions using \cref{fa} and setting $x_3=0$.  Right panel shows
    matrix elements $ c(\tau_{\mathrm{sink}}) $ obtained with
    different extraction methods.  }
  \label{fig-ratio-comp}
\end{figure}

We simultaneously fit all parameters according to the coefficient matrix
$M$.  This is not trivial but has the advantage that rows with a poor
signal are stabilized by others.  The fit parameter $x_3$ in
\cref{fa} is only resolvable if multiple $\tau_{\mathrm{sink}}$ per
ensemble are available.  Therefore, we perform in general a 3-exponent
fit where we set $x_3=0$.  To check whether this is justified or not we utilize
ensemble VIII which has three different source sink
separations $\tau_{\mathrm{sink}}/a = 9, 12, 15$, see
\cref{fig-ratio-comp}.
\newpage
In the right panel of \cref{fig-ratio-comp} we compare different
values for the matrix element as a function of 
$\tau_{\mathrm{sink}}$ and as a function of different extraction methods. 
More specifically, we compare the ratio method and the 3 exponent
excited-state fit to the full excited-state fit with 4 exponents which
is shown as a gray band.

We conclude from the right panel that at given statistics a
distinction between the different fit methods is not possible if the
source sink separation is $ \tau_{\mathrm{sink}}/a = 15$ which is
about $1\mathrm{fm}$.  This is consistent with what we have found in
\cite{Bali:2014nma}.  In this work we use the 3-exponent fit.

\section{Extraction of $A^{u-d}_{2,0}$, $B^{u-d}_{2,0}$ and $J^{u-d}$}
We utilize Baryon Chiral Perturbation Theory to extract $J^ {u-d}$ at
physical pion mass.  Our aim is to quote an upper and lower bound of
$A^{u-d}_{2,0}$, $B^{u-d}_{2,0}$ and $J^{u-d}$ including the
statistical error and the systematical error from the fit range
choice.  To that end we perform global combined fits to our lattice
ensembles with fit functions
$A^{u-d}_{2,0}(t, m_\pi^2, \vec{\Theta}_A)$ and
$B^{u-d}_{2,0}(t, m_\pi^2, \vec{\Theta}_B)$ known from (BChPT)
\cite{Wein:2014wma}. More specifically,
\begin{align}
  \label{eq:chia}
  A^{u-d}_{2,0}(t, m_\pi) \, &= \, \left[ 1 - \frac{(1 + 3\,g_A^2) \, m_\pi^2 \, 
                               \log( \frac{m_\pi^2}{\mu^2}) }{16\, f_\pi^2\, \pi^2} \right] \, L \, + m_\pi^2 
                               \, M_2^A \, + m_\pi^3\,M_3^A\,+ t(\, T_0^A + m_\pi^2 \, T_1^A)  \\
  \label{eq:chib}
  B^{u-d}_{2,0}(t, m_\pi) \, &= \, \frac{ g_A^2 \,  m_\pi^2\, \log(\frac{m_\pi^2}{ \mu^2}) }{16\, f_\pi^2 \,
                               \pi^2}\,L + \left[ 1 - \frac{(1 + 2 \,  g_A^2) \,m_\pi^2 \log(\frac{m_\pi^2}{\mu^2})}{16 \, f_\pi^2 \, \pi^2}\right]	\,L^B \, +m_\pi^2 \, M_2^B  + \, t(\, T_0^B + m_\pi^2 \, T_1^B)  .
\end{align}
with fit parameters $\vec{\Theta}_A = (L, M_2^A, M_3^A, T_0^A, T_1^A)$
and $\vec{\Theta}_B = (L, L^B, M_2^B, T_0^B, T_1^B)$.  The fit
parameters $T_1^A$ and $T_1^B$ are introduced by hand but they
naturally appear in next order of BChPT.  We do this since we consider
relatively large virtualities with
$|t_{\mathrm{max}} | = 0.6 \, \mathrm{GeV} \gg m_{\pi}$.

In order to get an upper and lower bound for $A^{u-d}_{2,0}$ and
$B^{u-d}_{2,0}$ we randomly sample fit range choices (to estimate the
systematical error).  For each sample we fit the bootstrap ensemble
(to estimate the statistical error).  Then we generate the
distributions of the fit parameters $\vec{\Theta}_A$ and
$\vec{\Theta}_B$ which contain the combined error.  In the last step
we draw fit parameters from the distributions $\vec{\Theta}_A$ and
$\vec{\Theta}_B$ and evaluate \cref{eq:chia} and \cref{eq:chib} with
$t = 0 \, \mathrm{GeV}$ and fixed pion mass $m_\pi$.  This yields
distributions for $A^{u-d}_{2,0}(t=0, m_\pi)$,
$B^{u-d}_{2,0}(t=0, m_\pi)$ and
$J^{u-d}(m_\pi)=(A^{u-d}_{2,0}({t=0, m_\pi}) + B^{u-d}_{2,0}({t=0,
  m_\pi}))/2$.  

\begin{figure}[!htb]
  \centering
  \includegraphics[width=0.98\textwidth]{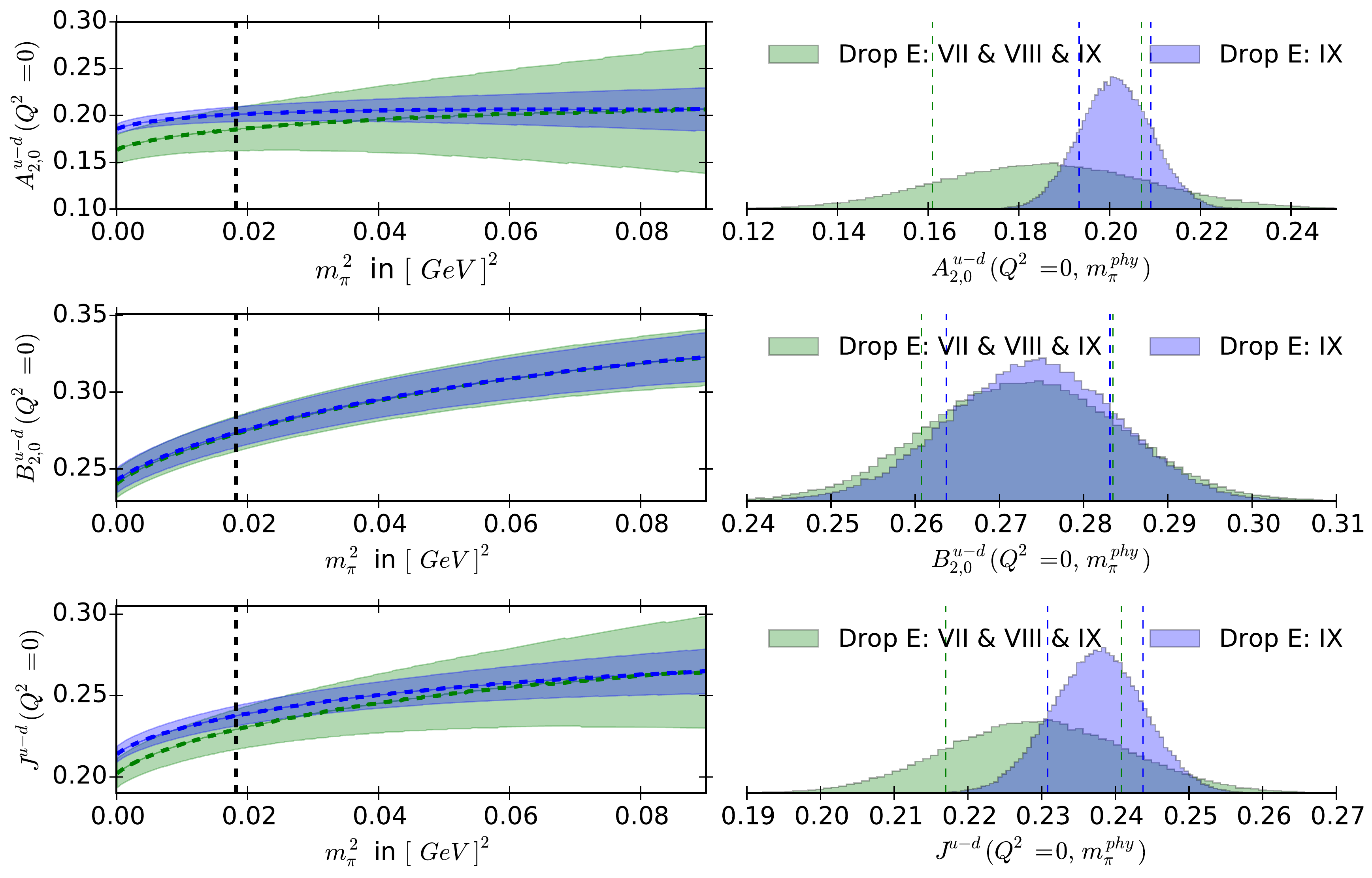}
  \caption{ Results for $A_{2,0}^{u-d}$ and $B_{2,0}^{u-d}$ and
    $J^{u-d}$.  The dashed black line indicates the physical pion
    mass.  We always exclude the heaviest ensemble IX due to the
    applicability of BChPT (blue).  To probe the stability of our
    Ansatz, we also show results where we additionally drop the two
    lightest ensembles VII \& VIII (green).  }
  \label{asdf}
\end{figure}

We define the lower and upper error as 0.16 and 0.84
quantiles respectively.  By repeating this process for many pion
masses $m_\pi$ we are able to draw error bands as shown in the left
panel of \cref{asdf} while the right panel shows the outcome for
$m_\pi = m_\pi^{\mathrm{phy}}$.
\section{Lattice results for the Axial-GFFs $\tilde{A}_{2,0}^{u-d}$ and
  $\tilde{B}_{2,0}^{u-d}$}
\begin{figure}[!htb]
  \centering
  \includegraphics[width=1.0\textwidth]{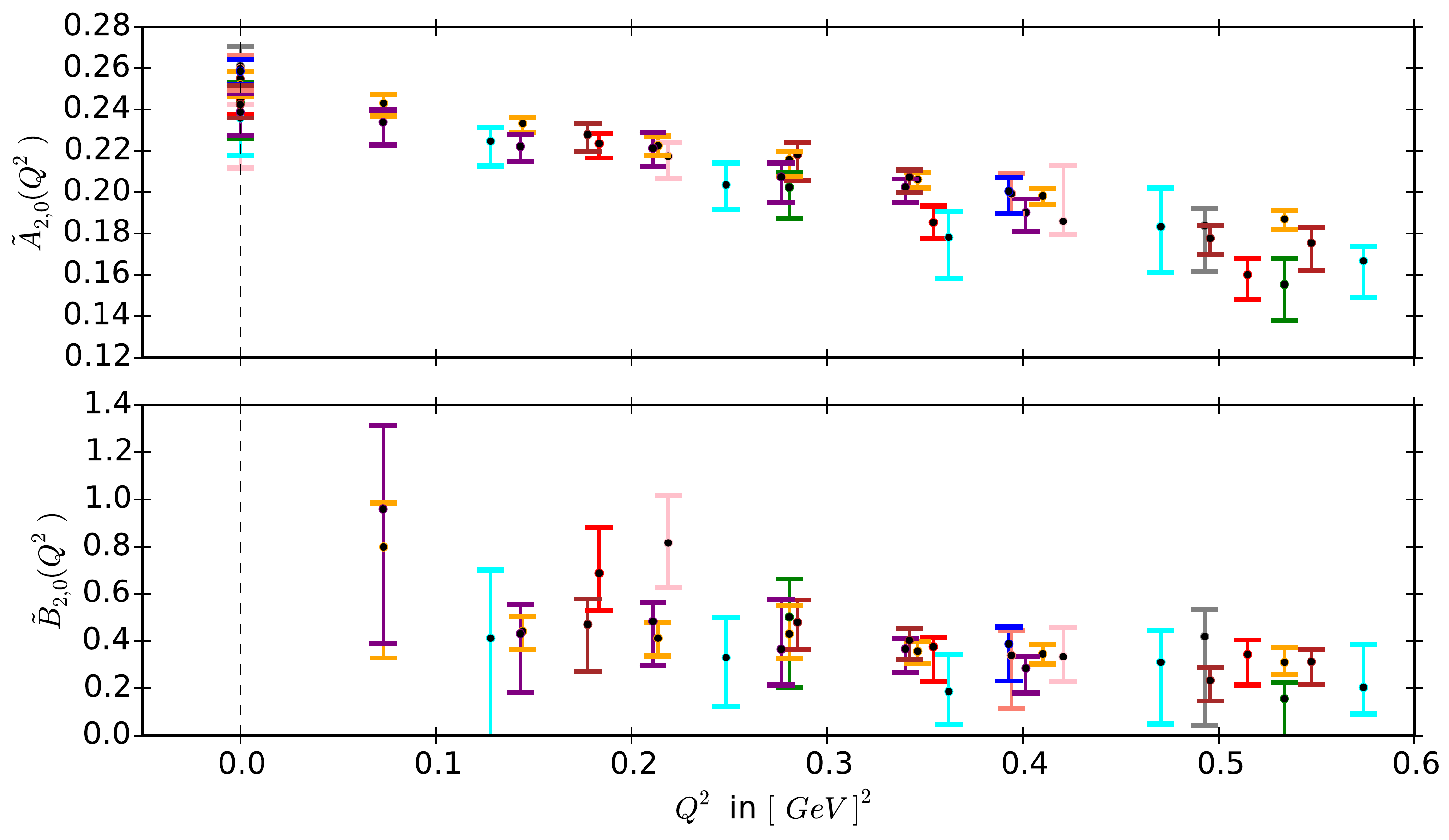}
  \caption{ Lattice results for $\tilde{A}_{2,0}^{u-d}$ and
    $\tilde{B}_{2,0}^{u-d}$.  For both GFFs we do not see a strong
    volume or pion mass dependence.  Most notably are the small errors
    of ensemble VI with a $64^3\times64$ volume and $Lm_\pi = 6.71$.
  }
  \label{lat-res-axial}
\end{figure}

We carefully study fit range dependencies (induced by the combined fit
to two- and three-point functions).  The error bars in
\cref{lat-res-axial} show the (statistic and systematic) one-sigma
error.  Each single error bar is constructed from a histogram with
sample-size s = (\# of fit range combinations) $\cdot$ (\#number of
bootstrap ensembles).  Here we have used $s=75\cdot300=22500$.  A more
detailed analysis will be published in the near future.

\section{Conclusion}
We determine $A_{2,0}^{u-d}$, $B_{2,0}^{u-d}$ and $J^{u-d}$ by virtue
of lattice QCD.  Further, utilizing Baryon Chiral Perturbation Theory,
we calculate the observables at the physical pion mass, where, we
focused on the construction of reliable errors.  This is achieved by
the histogram method which propagates the statical error and the
systematic error of the fit range choice. The final values are shown
in \cref{qwert}.
\begin{table}[!htbp]
  \centering
  \begin{tabular}{@{}llr@{}}
    \toprule
    Observable                                & $\overline{\mathrm{MS}} \ \mu=2 \mathrm{GeV}$ & \\ \midrule
    $A_{2,0}^{u-d}(Q^2 = 0, \ m_{\pi}^{phy})$ & 0.200 (-7/+9)                                 & \\
    $B_{2,0}^{u-d}(Q^2 = 0, \ m_{\pi}^{phy})$ & 0.275 (-11/+8)                                & \\
    $J^{u-d}(m_{\pi}^{phy})$                  & 0.238 (-8/+5)                                 & \\ \bottomrule
  \end{tabular}
  \caption{Table of results:
    The numbers correspond to the blue distributions shown in \cref{asdf}.
    They have a systematic error from the application of 
    Baryon Chiral Perturbation Theory to our lattice ensembles.
  }
  \label{qwert}
\end{table}

Further, we determine $\tilde{A}_{2,0}^{u-d}$ and $\tilde{B}_{2,0}^{u-d}$
which are the GFFs corresponding to the Axial-GPD, however, more work
has to be done especially to extrapolate $\tilde{B}_{2,0}^{u-d}$ to
$Q^2=0$.
\\~\\
Acknowledgements
\\~\\
This work is funded by Deutsche Forschungsgemeinschaft (DFG) within
the transregional collaborative research centre 55 (SFB-TRR55).  We
acknowledge computer time on SuperMUC at the Leibniz
Rechenzentrum in Garching.

\bibliography{roedl}

\begin{thebibliography}{1}

\bibitem{Mueller:1998fv}
D.~Müller, D.~Robaschik, B.~Geyer, F.~M. Dittes, and J.~Hořejši, ``{Wave
  functions, evolution equations and evolution kernels from light ray operators
  of QCD},'' {\em Fortsch. Phys.}, vol.~42, pp.~101--141, 1994.

\bibitem{Ji:1996ek}
X.-D. Ji, ``{Gauge-Invariant Decomposition of Nucleon Spin},'' {\em Phys. Rev.
  Lett.}, vol.~78, pp.~610--613, 1997.

\bibitem{Ji:1998pc}
X.-D. Ji, ``{Off forward parton distributions},'' {\em J. Phys.}, vol.~G24,
  pp.~1181--1205, 1998.

\bibitem{Ji:1996nm}
X.-D. Ji, ``{Deeply virtual Compton scattering},'' {\em Phys. Rev.}, vol.~D55,
  pp.~7114--7125, 1997.

\bibitem{LHPC:2003aa}
LHPC, P.~Hagler, J.~W. Negele, D.~B. Renner, W.~Schroers, T.~Lippert, and
  K.~Schilling, ``{Transverse structure of nucleon parton distributions from
  lattice QCD},'' {\em Phys. Rev. Lett.}, vol.~93, p.~112001, 2004.

\bibitem{Diehl:2003ny}
M.~Diehl, ``{Generalized parton distributions},'' {\em Phys. Rept.}, vol.~388,
  pp.~41--277, 2003.

\bibitem{Bali:2012qs}
G.~S. Bali {\em et~al.}, ``{Nucleon mass and sigma term from lattice QCD with
  two light fermion flavors},'' {\em Nucl. Phys.}, vol.~B866, pp.~1--25, 2013.

\bibitem{Bali:2014nma}
G.~S. Bali, S.~Collins, B.~Glässle, M.~Göckeler, J.~Najjar, R.~H. Rödl,
  A.~Schäfer, R.~W. Schiel, W.~Söldner, and A.~Sternbeck, ``{Nucleon
  isovector couplings from $N_f=2$ lattice QCD},'' {\em Phys. Rev.}, vol.~D91,
  no.~5, p.~054501, 2015.

\bibitem{Wein:2014wma}
P.~Wein, P.~C. Bruns, and A.~Schäfer, ``{First moments of nucleon generalized
  parton distributions in chiral perturbation theory at full one-loop order},''
  {\em Phys. Rev.}, vol.~D89, no.~11, p.~116002, 2014.

\end{thebibliography}
\bibliographystyle{ieeetr}
\end{document}